\documentclass[11pt]{article}
\usepackage{jheppub}
\usepackage{epsfig,amsfonts,amsthm}
\usepackage{amsmath,amssymb}
\usepackage[utf8]{inputenc}
\usepackage{float}
\usepackage{relsize}

\renewcommand{\Re}{\mathrm{Re }}
\renewcommand{\Im}{\mathrm{Im }}
\newcommand{\doublet}[2]{ \left( \begin{array}{c}#1 \\ #2 \end{array}\right) }

\newcommand{\lr}[1]{ \langle #1 \rangle}

\newcommand{\Z}{\mathbb{Z}}

\usepackage[usenames,dvipsnames]{xcolor}

\def\lsim{\mathrel{\rlap{\lower4pt\hbox{\hskip1pt$\sim$}}
    \raise1pt\hbox{$<$}}}         
\def\gsim{\mathrel{\rlap{\lower4pt\hbox{\hskip1pt$\sim$}}
    \raise1pt\hbox{$>$}}}         

	\title{
		Yet another lesson on the stability conditions in multi-Higgs potentials}

\author{Igor P. Ivanov,}
\author{Francisco Vaz\~{a}o}
\affiliation{Centro de Física Teórica de Particulas, Departamento de Física, Instituto Superior Técnico, Universidade de Lisboa, Lisboa}

\emailAdd{igor.ivanov@tecnico.ulisboa.pt} \emailAdd{francisco.vazao@tecnico.ulisboa.pt}

\abstract{
We discuss a rather common but often unnoticed pitfall which arises when deriving the bounded-from-below (BFB) conditions 
in multi-Higgs models with softly broken global symmetries.
Namely, necessary and sufficient BFB conditions derived for the case with an exact symmetry
can be ruined by introducing {\em soft} symmetry breaking terms.
Using $S_4$ and $A_4$-symmetric three-Higgs-doublet models as an example,
we argue that all published necessary and sufficient BFB conditions, 
even those which are correct for the exactly symmetric case, 
are no longer sufficient if soft symmetry breaking is added.
Using the geometric formalism, we derive the exact necessary and sufficient BFB conditions
for the 3HDM with the symmetry group $S_4$, either exact or softly broken, 
and review the situation for the $A_4$-symmetric case.
}
	
\begin{document}
	\maketitle

\section{Introduction}

Theoretical search for New Physics beyond the Standard Model (SM) is driven, in the absence of direct experimental indications, by ``educated guesses''. Theorists introduce a set of new fields, construct their interaction lagrangian based on a desired set of symmetries or anticipated phenomenological features, and then calculate observable consequences. New free parameters entering the lagrangian are either fixed by fitting the data or scanned over in search for viable and interesting versions of the model.

Before this phenomenological study begins, one must make sure that the model is mathematically self-consistent.
Going right to the point, let us consider a conservative but very rich and popular class of New Physics models, 
the $N$-Higgs-doublet model (NHDM), see e.g. reviews \cite{Branco:2011iw,Ivanov:2017dad} and references therein.
Its core element, the Higgs potential, must be bounded from below (BFB), at least at tree level, 
in order for a vacuum to exist. This implies that the coefficients of the quartic part of the potential
must satisfy certain inequalities known as stability, or BFB, conditions.
In simple cases, these can be written right away \cite{Branco:2011iw}; 
in other cases, one needs to resort to more elaborate tools such as copositivity methods \cite{Kannike:2012pe,Kannike:2016fmd}
or algebraic and geometric constructions in the bilinear space \cite{Nagel:PhD,Ivanov:2005hg,Nishi:2006tg,Maniatis:2006fs,Ivanov:2006yq,Nishi:2007nh,Ivanov:2010ww,Ivanov:2010wz,Maniatis:2014oza,Maniatis:2015gma}. 
Thanks to such methods, there now exists the set of necessary and sufficient BFB conditions for the general 2HDM \cite{Ivanov:2006yq}.
For three Higgs doublets, the problem remains unsolved in its full generality.
However, in the particular versions of the 3HDM equipped with additional global symmetries,
the structure of the Higgs potential simplifies, and one can (hope to) obtain such BFB conditions.

It turns out that this task contains pitfalls. One of them was recently pointed out in \cite{Faro:2019vcd}:
even if one has a potential with a valid neutral minimum and wants to constrain the quartic interactions via the BFB conditions, 
it is imperative to check stability along all directions in the Higgs space including the charge-breaking\footnote{That is, 
if a minimum happened to lie along such a direction, it would be a charge-breaking vacuum.} directions.
By neglecting this requirement, one may overlook a ``hidden pathology'' of a model.
Ref.~\cite{Faro:2019vcd} illustrated this pitfall with a 3HDM potential with symmetry group $U(1)\times U(1)$,
which has a normally looking neutral vacuum, with all scalar masses squared positive, 
which even appears to satisfy BFB constraints if one explores all neutral directions.
Nevertheless, the potential is unbounded from below along certain charge breaking directions.

In the present paper, we discuss another pitfall, which may seem surprising and which is applicable 
to models with softly broken global symmetries.
Suppose one has derived necessary and sufficient BFB conditions for a multi-Higgs model with an exact global symmetry group.
Then, by introducing {\em soft} symmetry breaking terms, one can inadvertently render these BFB conditions insufficient.
We will explain the origin of this baffling phenomenon and, taking it into account, derive the exact necessary and sufficient conditions
for the $S_4$-symmetric 3HDM. 
On the way, we will also comment on the validity of several BFB results for the $A_4$-symmetric 3HDM scattered across the literature.

Before going into details, let us address the question of whether these mathematical intricacies are unavoidable 
when undertaking a phenomenology-focused multi-Higgs study.
Certainly, conditions which are necessary but not sufficient are dangerous, as one may be led to explore in detail
a model which is self-inconsistent. On the other hand, finding sufficient but not necessary BFB conditions is safe,
in the sense that if a model passes them, it can be used for a phenomenological study.
Since finding a set of sufficient conditions is rather easy, it has become a standard approach 
when performing numerical scans in the parameter space of models based on elaborate scalar potentials.
A drawback of this approach is that one over-restricts the model.
Indeed, focusing only on those models where the (easy) sufficient BFB conditions apply, 
one can miss whole regions in the parameter space which are perfectly BFB and may even lead 
to unique, intriguing phenomenology.
Thus, these intricacies become unavoidable if one aims to give a systematic analysis in a class of multi-Higgs models.

The outline of the paper is as follows.
In the next section we will set up the notation, formulate the problem for the $A_4$-symmetric 3HDM,
and list BFB conditions stated in several publications.
In section~\ref{section-U1U1} we will describe a geometric approach to establishing the BFB conditions. 
It is here where we explain
why soft breaking terms can render the BFB conditions insufficient.
Section~\ref{section-S4} covers the $S_4$-symmetric case, which is the $A_4$ 3HDM with all coefficients real.
We first report the derivation of the exact necessary and sufficient conditions 
and then comment on several publications which mentioned the BFB problem.
The status of the BFB conditions in the full $A_4$ 3HDM is reviewed in section~\ref{section-A4}. 
In the last section, we draw our conclusions.
Several appendices provide details of the calculations in support of statements made in the main text.

\section{The $A_4$ 3HDM scalar potential}

\subsection{Two parametrizations}

Using three Higgs doublets with equal electroweak quantum numbers $\phi_i$, $i = 1, 2, 3$, 
one can construct only one $A_4$-invariant quadratic term and five $A_4$-invariant quartic terms for the Higgs potential
\cite{Ma:2001dn,Ishimori:2010au,Ivanov:2012ry,Ivanov:2012fp}. 
Its traditional form, in the notation of \cite{Toorop:2010ex}, is
\begin{eqnarray}
V &=& \mu^2(\phi_1^\dagger\phi_1 + \phi_2^\dagger\phi_2 + \phi_3^\dagger\phi_3) + 
\lambda_{1}(\phi_1^\dagger\phi_1 + \phi_2^\dagger\phi_2 + \phi_3^\dagger\phi_3)^2 \nonumber\\[2mm]
&& + \lambda_{3}\left[(\phi_1^\dagger\phi_1)(\phi_2^\dagger\phi_2) + (\phi_1^\dagger\phi_1)(\phi_3^\dagger\phi_3) + (\phi_2^\dagger\phi_2)(\phi_3^\dagger\phi_3)\right] + \lambda_{4} (|\phi_1^\dagger\phi_2|^2 + |\phi_2^\dagger\phi_3|^2 + |\phi_3^\dagger\phi_1|^2)\nonumber\\[2mm]
&& + \frac{\lambda_{5} }{2} \left\{e^{i\epsilon}\left[(\phi_1^\dagger\phi_2)^2 + (\phi_2^\dagger\phi_3)^2 + 
(\phi_3^\dagger\phi_1)^2 \right] + H.c.\right\}\,.\label{VA4-1}
\end{eqnarray}
This potential is invariant under sign flips of individual doublets (the group $\Z_2 \times \Z_2$) 
and under cyclic permutations of the three doublets (the group $\Z_3$),
which form the global symmetry group $(\Z_2 \times \Z_2) \rtimes \Z_3 \simeq A_4$ of order 12.
In addition, the model contains 12 generalized $CP$ symmetry such as, for example,
$\phi_1 \mapsto \phi_2^*$, $\phi_2 \mapsto \phi_1^*$, $\phi_3 \to \phi_3^*$. 
Notice that the parameter $\epsilon$ is defined modulo $2\pi/3$. For example,
the model with $\epsilon = 2\pi/3$ can be brought to exactly the same model with $\epsilon = 0$
by discrete rephasing of the three doublets.

If $\sin\epsilon = 0$ (which, according to the above remark, is equivalent to $\sin 3\epsilon = 0$),
then the potential is invariant under the usual $CP$ transformation $\phi_i \mapsto \phi_i^*$
as well as arbitrary (not just cyclic) permutations of the three doublets.
Since $(\Z_2 \times \Z_2) \rtimes S_3 \simeq S_4$, we call this model $S_4$-symmetric 3HDM.
Finally, if $\lambda_5 = 0$, the potential becomes invariant under arbitrary rephasing of the three doublets,
and the symmetry group is $[U(1)\times U(1)] \rtimes S_3$.
We will refer to this model as a rephasing invariant model.

It turns out that the 3HDM with an exact $A_4$ symmetry and no other fields 
is too rigid to be suitable for phenomenological applications. 
Extended to the Yukawa sector, it does not allow one to obtain a realistic quark sector,
see an exhaustive census of cases in \cite{Felipe:2013ie,Felipe:2013vwa} 
and the fundamental origin of this obstacle in \cite{Felipe:2014zka}.
A popular way to make use of the structures imposed by the symmetry is to allow for soft symmetry breaking,
that is, to complement the potential with various quadratic terms violating $A_4$.
Soft breaking terms relax the rigid structure of the discrete symmetry group
and affect observables such as Higgs masses, but the quartic potential remains unchanged.
Thus, if one obtains the exact necessary and sufficient BFB conditions for the case with an exact symmetry,
one could expect to be able to use them for the softly broken versions as well.

An alternative form of potential \eqref{VA4-1} was used in \cite{Degee:2012sk}:
\begin{eqnarray}
V&=&-\frac{M_0}{\sqrt{3}}\left(\phi_1^{\dagger}\phi_1+\phi_2^{\dagger}\phi_2+\phi_3^{\dagger}\phi_3\right)+\frac{\Lambda_0}{3}\left(\phi_1^{\dagger}\phi_1+\phi_2^{\dagger}\phi_2+\phi_3^{\dagger}\phi_3\right)^2\nonumber\\ 
&&+\frac{\Lambda_3}{3}\left[(\phi_1^{\dagger}\phi_1)^2+(\phi_2^{\dagger}\phi_2)^2+(\phi_3^{\dagger}\phi_3)^2-(\phi_1^{\dagger}\phi_1)(\phi_2^{\dagger}\phi_2)-(\phi_2^{\dagger}\phi_2)(\phi_3^{\dagger}\phi_3)-(\phi_3^{\dagger}\phi_3)(\phi_1^{\dagger}\phi_1)\right]\nonumber\\
&&+\Lambda_1\left[(\Re\phi_1^{\dagger}\phi_2)^2+(\Re\phi_2^{\dagger}\phi_3)^2+(\Re\phi_3^{\dagger}\phi_1)^2\right]\nonumber\\
&&+\Lambda_2\left[(\Im\phi_1^{\dagger}\phi_2)^2+(\Im\phi_2^{\dagger}\phi_3)^2+(\Im\phi_3^{\dagger}\phi_1)^2\right]\nonumber \\
&&+\Lambda_4\left[(\Re\phi_1^{\dagger}\phi_2)(\Im\phi_1^{\dagger}\phi_2)+(\Re\phi_2^{\dagger}\phi_3)(\Im\phi_2^{\dagger}\phi_3)+
(\Re\phi_3^{\dagger}\phi_1)(\Im\phi_3^{\dagger}\phi_1)\right]\,,\label{VA4-2}
\end{eqnarray}
with all coefficients being real.
The motivation behind this parametrization is in its connection with the bilinear formalism, which can be advantageous
for certain problems. For completeness, we provide in Appendix~\ref{appendix:bilinear} more details on this relation.
The two sets of parameters are related by
\begin{equation}
-{M_0 \over \sqrt{3}} = \mu^2\,,\quad \Lambda_0 = 3\lambda_1 + \lambda_3 \,,\quad 
\Lambda_3 = - \lambda_3\,,\quad \Lambda_{1,2} = \lambda_4 \pm \lambda_5 \cos\epsilon\,,\quad 
\Lambda_4 = -2\lambda_5 \sin \epsilon\,.
\end{equation}
By setting $\Lambda_4 = 0$, one obtains the $S_4$ 3HDM. 
If, in addition, one sets $\Lambda_1 = \Lambda_2$, one arrives 
at the rephasing invariant model.

\subsection{Linking the BFB conditions with minimization} 

Following \cite{Degee:2012sk}, we further geometrize the model by defining 
$r_0 = (\phi_1^{\dagger}\phi_1+\phi_2^{\dagger}\phi_2+\phi_3^{\dagger}\phi_3)/\sqrt{3}$ 
and expressing the potential as
\begin{equation}
V = -M_{0}r_{0}+r_{0}^2 v_4\,,\quad 
\mbox{where}\quad v_4 = \Lambda_{0} + \Lambda_{1} x + \Lambda_2 y + \Lambda_3 z + \Lambda_4 t\,.\label{VA4-3}
\end{equation}
The definition of the dimensionless real variables $x, y, z, t$ is immediately read off Eq.~\eqref{VA4-2}.
These real variables are not independent. They fill a bounded region in $\mathbb{R}^4$
which is called the orbit space. 
This orbit space was the key object in Ref.~\cite{Degee:2012sk} and it is also the main tool for derivation of the BFB conditions
in the present paper.

The quadratic part of this model is extremely simple due to the group-theoretic arguments:
if all Higgs doublets transform as an irreducible representation of the chosen symmetry group, 
one can construct only one quadratic group invariant respecting gauge symmetries.
This extreme simplicity, in turn, leads to a natural separation of the ``angular'' and ``radial''
problems when looking for the global minimum of the potential \eqref{VA4-3}.
It also links this task with the search for the BFB conditions. 
At first, one fixes $r_0$ and minimizes $v_4$ over the orbit space. This minimal value $v_{4, min}$ must be non-negative\footnote{In fact, it must be strictly positive.
Flat directions are not allowed for the quartic potential because the quadratic term is definitely negative along any direction.
However, we stay with the term ``non-negative'' to cover in future the cases of soft symmetry breaking,
where the modified quadratic term relaxes this requirement and, in principle, allows for flat directions
of the quartic potential.};
otherwise, the quartic potential would go to minus infinity at asymptotically large values of the fields.
Once $v_{4, min}$ is found, one solves the ``radial'' problem, finds $r_0$, and computes the value of the
potential at the global minimum:
\begin{equation}
V_{min} = -\frac{M_0^2}{4v_{4, min}}\,.
\end{equation}
The minimum value of the potential is attained at the point of the smallest $v_4$.

This relation between the two problems, which hinges on the simplicity of the quadratic term,
allows us to look at the BFB task from a slightly different perspective.
Instead of striving for infinitely large values of Higgs fields and checking the positivity of the quartic potential 
along all directions, one can look back at the {\em full} potential and identify all Higgs field configurations
which can, in principle, become the global minimum of the potential for some values of the coefficients.
These directions correspond to some points in the orbit space.
When all of them are found, one can just require that $v_4 \ge 0$ at all such points. 
This set of conditions will yield the necessary and sufficient BFB constraints for our model.

However, this direct relation between the two problems disappears once soft breaking terms are allowed.
The correct procedure for the BFB problem is just to keep the same expressions as in the model with an exact symmetry.
Relying on the knowledge of a minimum in the model with a softly broken symmetry --- a tacit assumption in virtually all previous works 
--- may lead to pathological cases with nice minima but unbounded potentials. 

\subsection{The literature on the BFB conditions in $A_4$ 3HDM} \label{literature_conditions}

It is among the goals of the present work to revisit what is claimed in the literature on the BFB conditions in $A_4$ 3HDM.
The symmetry group $A_4$ is a frequent guest in New Physics models, especially in the context of the flavor puzzle
and neutrino sector, see e.g. \cite{Ishimori:2010au,King:2013eh} and references therein. 
It is often put in the context of three Higgs doublets.
As a result, there is quite a number of papers which deal with the 3HDM potential with the (softly broken) $A_4$ symmetry group
and, in principle, need to assure that the potential is bounded from below. 

Here, we collect the BFB conditions for this model explicitly quoted in literature. 
In all cases, the conditions were recast in the notation of Eq.~\eqref{VA4-1}.
It turns out that not all papers on this model deal with the BFB conditions.
For example, the influential Ref.~\cite{Ma:2001dn} did not address this issue and immediately proceeded to the phenomenology.
Also, in many cases, the language used by the authors does not make it clear if the conditions are claimed to be just necessary, just sufficient, 
or necessary and sufficient. 

In all cases, one straightforward condition was indicated, $\lambda_1 \geq 0$, and the difference 
was only in the supplementary requirements, which we now list:
\begin{itemize}
    \item Reference \cite{Toorop:2010ex}:  $\lambda_1 + \lambda_3 + \lambda_4 +\lambda_5 \cos{\epsilon} \geq 0$.
    \item Reference \cite{Dekens:2011}: several necessary conditions were obtained by checking various directions 
    in the Higgs space:
    \begin{equation}
    3\lambda_1 + \lambda_3 + \lambda_4 +\lambda_5 \cos{\epsilon} \geq 0\,, \quad 
    3\lambda_1 + \lambda_3 + \lambda_4 \geq 0\,, \quad 
    4\lambda_1 + \lambda_3 \geq 0\,, \quad 
	4\lambda_1 + \lambda_3 + \lambda_4 - |\lambda_5| \geq 0\,.\label{Dekens}
    \end{equation}
	\item Reference \cite{Boucenna:2011tj}: an elaborate scalar sector was assumed, which was constrained 
	by a rather sophisticated overall condition on 15 coefficients of the potential. 
	Focusing on the pure $A_4$ 3HDM part, one can translate this constraint into two expressions:
	$3\lambda_1 + \lambda_3 \geq 0$ and $3\lambda_1 + \lambda_3 + \lambda_4 -|\lambda_5| \geq 0$. 
    \item Reference \cite{Pramanick:2017wry}: 
    for the $S_4$ symmetric case, two conditions were given:
    \begin{equation}
    3\lambda_1 + \lambda_3 + \lambda_4+ \lambda_5 \geq 0\,, \quad  
    3\lambda_1 + \lambda_3 + \lambda_4 - \lambda_5/2 \geq 0\,.\label{Pramanick-S4}
    \end{equation}
    For the $A_4$ case, which was considered in the appendix, no additional BFB condition was explicitly stated, 
    although the existence of one solution required that 
    \begin{equation}
	4\lambda_1 + \lambda_3 + \lambda_4 - |\lambda_5| \geq 0\,.\label{Pramanick-A4-extra}
	\end{equation}
    \item Reference \cite{Chakrabarty:2018yoy}: two of their conditions can be compactly written as 
    $\lambda_1 - |\lambda_4| +\lambda_5 \cos{\epsilon} \geq 0$, while yet another one (assuming it contains a misprint to be corrected)
    leads to $3\lambda_1 + \lambda_3 + \lambda_4 +\lambda_5 \geq 0$.
\end{itemize}

\subsection{A quick check for the rephasing invariant model}

Before we go into our own derivation of the BFB conditions for the $S_4$ and $A_4$ models,
it is useful to perform a quick check of the validity of the conditions stated above in the simpler case of the rephasing
invariant model with $\lambda_5 = 0$.
Since such a potential acquires the symmetry group $U(1)\times U(1)$, the results of the recent paper \cite{Faro:2019vcd} apply. 
In Appendix~\ref{appendix:checks}, we follow the algorithm presented there to derive the necessary and sufficient BFB conditions for this case: 
\begin{eqnarray}
\mbox{neutral:}&\qquad& \lambda_1 \ge 0\,, \quad
3\lambda_1 + \lambda_3 + \lambda_4 \ge 0\,, \label{final-U1U1-neutral}\\
\mbox{charge-breaking:}&\qquad& 4\lambda_1 + \lambda_3 \ge 0\,, \quad
3\lambda_1 + \lambda_3 + {1\over 4}\lambda_4 \ge 0\,.\label{final-U1U1-CB}
\end{eqnarray}
We explicitly indicated here the conditions which come from checking the neutral and charge-breaking directions in the Higgs doublet space.
We stress that checking the BFB conditions along charge-breaking directions is mandatory even if the potential
has a normally looking neutral minimum \cite{Faro:2019vcd}.

Comparison of the above listed results of papers 
\cite{Toorop:2010ex,Pramanick:2017wry,Chakrabarty:2018yoy,Dekens:2011,Boucenna:2011tj} at $\lambda_5 = 0$ with these conditions shows that none of them gave the full list of necessary and sufficient BFB constraints even in this simple case.
More specifically, all papers apart from \cite{Toorop:2010ex} correctly found the two conditions \eqref{final-U1U1-neutral},
however none of them fully matches the charge-breaking part of our conditions \eqref{final-U1U1-CB}.
In particular, our last condition does not appear in any previous publication.

This is not surprising. In fact, most of the authors derived their BFB conditions under the assumption, implicit or explicit,
that only neutral directions should be checked, which, as we know after \cite{Faro:2019vcd}, is not correct in general.
The only exception is \cite{Dekens:2011} where at least one charge-breaking condition was correctly found
because the author explicitly checked such directions.

But does it mean all these previous works are {\em wrong} already for the rephasing invariant model?
In fact, no. We will show in the next section that the conditions of \cite{Dekens:2011,Pramanick:2017wry}
supplemented with the assumption of staying in a neutral minimum and with one more, missing step in the derivation,
are fully equivalent to our conditions \eqref{final-U1U1-neutral} and \eqref{final-U1U1-CB}. 
However, this equivalence holds only for the model with an {\em exact} rephasing symmetry.
Soft breaking terms will ruin it and render their conditions insufficient 
while keeping our Eqs.~\eqref{final-U1U1-neutral} and \eqref{final-U1U1-CB} unchanged.

In anticipation of what we will explain in the next section,
let us rewrite our conditions \eqref{final-U1U1-neutral} and \eqref{final-U1U1-CB} 
using the notation of \eqref{VA4-2}:
\begin{eqnarray}
\mbox{neutral:}&\qquad& \Lambda_0 + \Lambda_3 \ge 0\,, \quad
\Lambda_0 + \Lambda_1 \ge 0\,, \label{final-U1U1-neutral-Lam}\\
\mbox{charge-breaking:}&\qquad& \Lambda_0 + \frac{\Lambda_3}{4} \ge 0\,, \quad
\Lambda_0 + \frac{\Lambda_1}{4} \ge 0\,.\label{final-U1U1-CB-Lam}
\end{eqnarray}
The remarkable simplicity of these expressions indicates that the notation of \eqref{VA4-2} is indeed
most appropriate for their derivation. We are now ready to explain 
how Eqs.~\eqref{final-U1U1-neutral-Lam} and \eqref{final-U1U1-CB-Lam} can be obtained in a much more elegant way.

\section{BFB conditions from geometry}\label{section-U1U1}

\subsection{Geometric treatment of the rephasing invariant model}

\begin{figure} [ht]
	\begin{center}
		\includegraphics[height=5cm]{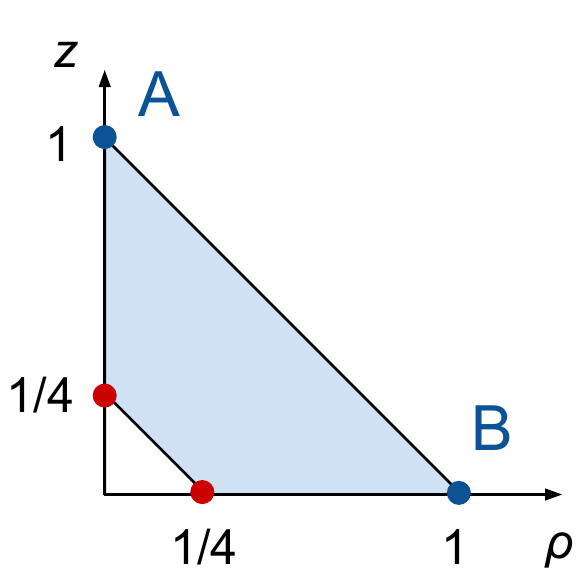}
		\caption{The orbit space on the $(\rho,z)$ plane of the rephasing-invariant version of the $A_4$ 3HDM.
		Two blue dots correspond to possible neutral minima, which the two read dots correspond to the two possible charge-breaking minima.}
		\label{fig-rho-z}
	\end{center}
\end{figure}

In order to clarify the situation, let us now rederive the BFB conditions for the rephasing invariant model
via the geometric method developed in \cite{Degee:2012sk}.
The $A_4$-symmetric potential \eqref{VA4-2} acquires the full rephasing symmetry if $\Lambda_4 = 0$ and $\Lambda_1=\Lambda_2$.
Thus, $v_4$ of Eq.~\eqref{VA4-3} depends now on two variables only:
\begin{equation}
v_4 = \Lambda_{0} + \Lambda_{1} \rho + \Lambda_3 z\,, \quad \rho \equiv x+y\,.\label{v4-U1U1}
\end{equation}
These two variables are bounded by the following linear relations \cite{Degee:2012sk,Ivanov:2010ww}:
\begin{equation}
\rho \ge 0\,, \quad z\ge 0\,,\quad 1 \ge \rho+z \ge 1/4\,.\label{relations-U1U1}
\end{equation}
Notice that neutral directions in the Higgs field space always correspond to $\rho + z = 1$, while charge-breaking directions
correspond to $\rho + z < 1$.
The resulting orbit space, that is, the space of all allowed Higgs field configurations squashed on this plane, 
has the trapezoidal shape shown in Fig.~\ref{fig-rho-z}.
It has four vertices. Two vertices are along neutral directions with the following representative vacuum expectation values (vev) alignments: 
\begin{eqnarray}
\mbox{point A:} &\quad& z=1,\, \rho = 0 \quad \Rightarrow \quad \lr{\phi_i^0} \propto (1, 0, 0),\nonumber\\
\mbox{point B:} &\quad& z=0,\, \rho = 1 \quad \Rightarrow \quad \lr{\phi_i^0} \propto (1, 1, 1),\label{minU1U1-AB}
\end{eqnarray}
The other two vertices, marked with red dots, correspond to charge-breaking directions with the following representative points:
\begin{eqnarray}
z=1/4,\, \rho = 0  \quad &\Rightarrow& \quad  \lr{\phi_i} \propto 
\doublet{0}{1},\ 
\doublet{1}{0},\
\doublet{0}{0},\nonumber\\[2mm]
z=0,\, \rho = 1/4  \quad &\Rightarrow&  \quad \lr{\phi_i} \propto 
\doublet{0}{1},\ 
\doublet{\sqrt{3}/2}{-1/2},\
\doublet{\sqrt{3}/2}{1/2}.
\label{minU1U1-charged}
\end{eqnarray}
All these representative vev alignments are defined up to an arbitrary rephasing and permutation.

The power of the geometric picture highlighted in \cite{Degee:2012sk} is that, just looking at it, we immediately conclude that, 
if we want to avoid massless scalars, the global minimum can only be at one of these four vertices.
Indeed, since the potential is now a {\em linear} function of the two variables,
the equipotential surfaces at any given $r_0$ are represented by straight lines orthogonal to the direction
of the ``steepest descent'' $\vec n = - \vec \Lambda \equiv -(\Lambda_1, \Lambda_3)$.
Depending on the values and signs of $\Lambda_1$ and $\Lambda_3$, one can easily locate the point of the entire orbit space
which is the farthest in the direction of the steepest descent.
It is either a vertex or, in exceptional cases, an entire segment linking two vertices. 
However in this latter case, there will be an entire manifold of vevs corresponding to the same depth of the potential,
which is only possible in the presence of an accidental continuous symmetry and the resulting massless Higgses.

This construction is immediately translated into the BFB conditions for this model.
Indeed, the BFB conditions are equivalent to the requirement that $v_4 \ge 0$ 
everywhere in the orbit space. However, one does not need to check {\em all} points of the orbit space.
The orbit space is a convex polygon and, therefore, one only needs to write these conditions
for its four vertices:
\begin{equation}
\Lambda_0 + \Lambda_3 \ge 0\,, \quad 
\Lambda_0 + \Lambda_1 \ge 0\,, \quad 
\Lambda_0 + \frac{\Lambda_3}{4} \ge 0\,, \quad 
\Lambda_0 + \frac{\Lambda_1}{4} \ge 0\,.\label{BFB-U1U1} 
\end{equation}
These necessary and sufficient conditions coincide with 
Eqs.~\eqref{final-U1U1-neutral-Lam} and \eqref{final-U1U1-CB-Lam}
which were derived in Appendix~\ref{appendix:checks} through a much more elaborate procedure.

Notice that none of these conditions can be thrown away.
Indeed, if $\Lambda_0 + \Lambda_1 \ge 0$, it may still happen that $\Lambda_0 < 0$, 
so that $\Lambda_0 + \Lambda_1/4$ can become negative, and the potential would be unbounded from below along some directions.
It is if and only if all four conditions are satisfied that the potential becomes truly bounded from below.

\subsection{Is checking neutral minima enough?}

In the case of the rephasing invariant version of the $A_4$ 3HDM, 
the (neutral) BFB conditions derived in \cite{Dekens:2011,Pramanick:2017wry} coincide
with the first two conditions of \eqref{BFB-U1U1}.
Clearly, they are not sufficient per se: one can satisfy them and yet violate the remaining two.
However the conditions of \cite{Dekens:2011,Pramanick:2017wry} can be made equivalent to ours
if we complement them with two extra arguments.

First, in the situation we consider, the two tasks --- finding the global minimum and establishing the BFB conditions ---
are intimately related. 
We want our minimum to be neutral, and this automatically gives priority to the first two conditions in \eqref{BFB-U1U1}.
It is possible to construct a version of the model for which checking the other two conditions would be needed.
But this can happen {\em only if we allow for a charge-breaking minimum}.
If we insist on working in a neutral minimum, then the first two conditions suffice.
The other two will be satisfied automatically.

Is it a trivial remark? Not at all. This conclusion hinges on the all-important fact
that the orbit space in our case is {\em convex}. Therefore, if a vertex of the orbit space corresponds to a minimum,
with all Higgs masses squared positive, then it is automatically the global minimum.
In particular, it removes the possibility of an even deeper charge-breaking vacuum or a direction of unbounded potential anywhere in the Higgs space.
In this version of $A_4$ 3HDM with the exact rephasing symmetry, 
there is no room for the pathological situation described in \cite{Faro:2019vcd}.

\begin{figure} [ht]
	\begin{center}
		\includegraphics[height=5cm]{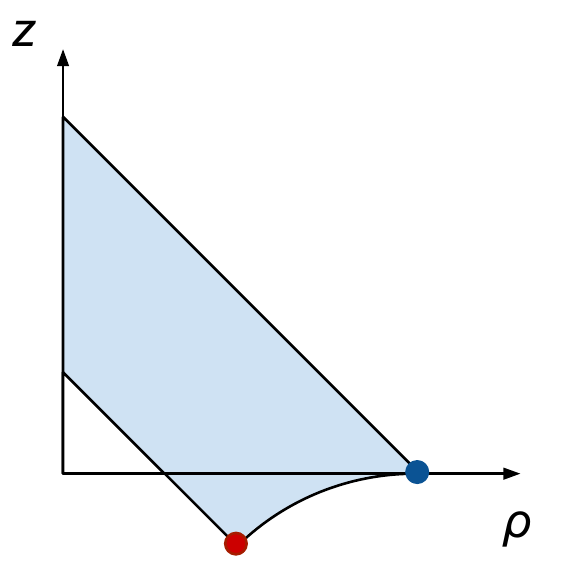}
		\caption{A hypothetical case of an orbit space lacking convexity
			in its charge-breaking part. }
		\label{fig-rho-z-2}
	\end{center}
\end{figure}

However, this feature holds in our case because we have explicitly constructed the orbit space.
It is conceivable that in other, more elaborate situations, the orbit space loses 
convexity in its charge-breaking part.
Imagine that the orbit space contains a ``protrusion'' as indicated in Fig.~\ref{fig-rho-z-2}.
Then, selecting the neutral vacuum (blue dot) and checking that charged Higgs masses squared are all positive
will be {\em insufficient} to assure that the chosen vacuum is global.
One can easily draw an equipotential surface (a straight line) through the blue point 
which would cut a part of the protrusion. This means that the charge-breaking cusp point 
marked with a red dot would lie even lower, and the potential may even be unbounded in the charge-breaking directions.
Thus, in that case, requiring that $v_4 \ge 0$ for all neutral vacua would be insufficient for the BFB conditions.
One would be forced to verify $v_4 \ge 0$ at the charge-breaking cusp point as well.

The lesson from this exercise is that the BFB conditions of \cite{Dekens:2011,Pramanick:2017wry},
computed for rephasing invariant version of the model, are not only necessary but also sufficient for any phenomenologically acceptable model.
However proving it requires an extra step just made --- verifying the convexity of the orbit space --- 
which was not present in \cite{Dekens:2011,Pramanick:2017wry}.

\subsection{Adding soft symmetry breaking}

Now we are ready to explain the feature announced in the introduction.
Suppose we add soft breaking terms to the potential and insist on having a neutral vacuum.
The quartic part of the potential does not change; therefore the full BFB conditions \eqref{BFB-U1U1}
remain as valid and complete as before.

What changes is the relation between charged Higgs masses and the directions of the steepest descent on the $(\rho,z)$ plane.
In the case of an exact symmetry, this plane represented the only degrees of freedom for the potential change
at any fixed $r_0$. 
With soft breaking terms, there is a new contribution not grasped by this plane.
In particular, it may easily happen that the quartic potential {\em decreases} 
as one ventures into the charge-breaking part of the orbit space, but, for small deviations and small $r_0$,
the overall potential is locally stabilized by the extra quadratic terms making charged Higgs masses squared positive.
It may be only at larger values of $r_0$ that a deeper charge-breaking minimum or a unbounded direction appears.

Thus, we come to the conclusion that, in the case of softly broken symmetry, the conditions derived in 
\cite{Dekens:2011,Pramanick:2017wry} are no longer sufficient. 
Therefore, to eliminate this danger, we must impose all four inequalities \eqref{BFB-U1U1}.

\section{The $S_4$-symmetric case} \label{section-S4}

\subsection{The case with an exact $S_4$ symmetry}

\begin{figure} [ht]
	\centering
	\includegraphics[height=4cm]{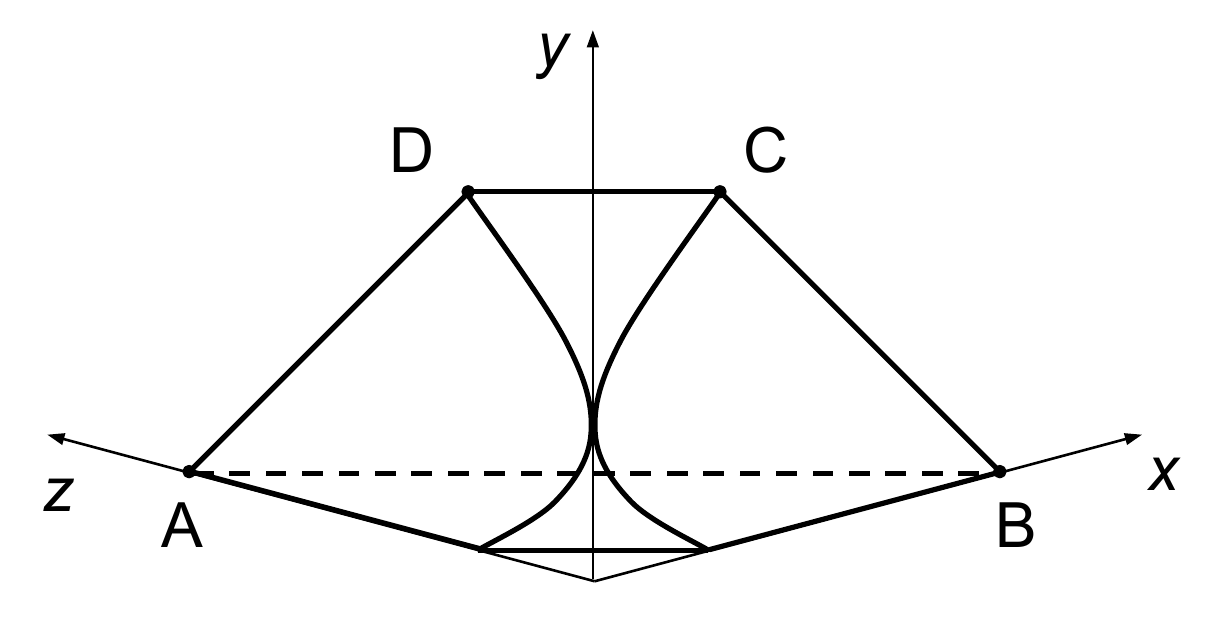}\hspace{5mm}
	\includegraphics[height=4cm]{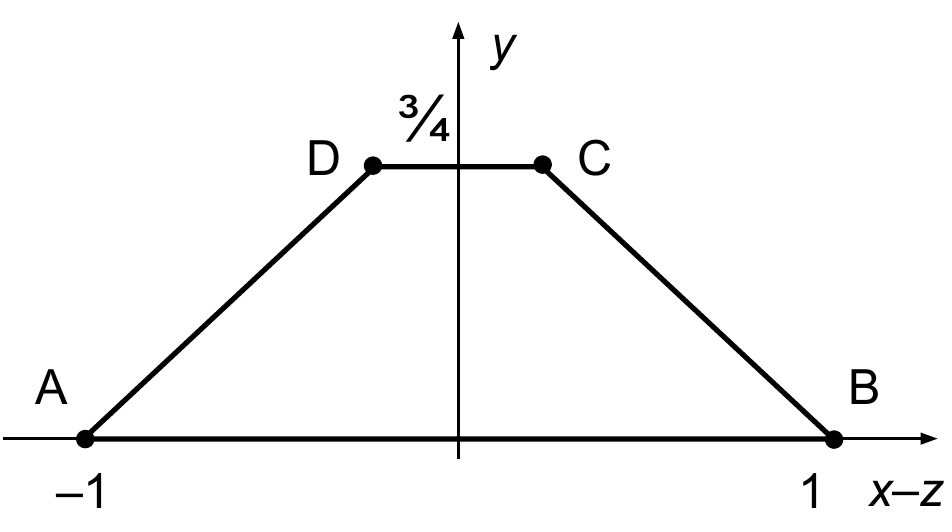}
	\caption{Left: a sketch of the orbit space of the $S_4$-symmetric 3HDM in the $(x,y,z)$-space. Right: the neutral orbit space
		in the $(x-z,y)$ plane. On each plot, the four dots $A$, $B$, $C$, and $D$ mark the positions of the possible neutral global minima.}
	\label{fig-S4-orbit}
\end{figure}

Ler us now apply the same geometric approach to the $S_4$ symmetric 3HDM.
Its scalar potential is given by \eqref{VA4-1} with $\sin\epsilon = 0$,
or equivalently by \eqref{VA4-2} with $\Lambda_4 = 0$.
Its $v_4$ is
\begin{equation}
v_4 = \Lambda_{0} + \Lambda_{1} x + \Lambda_2 y + \Lambda_3 z\,.\label{v4-S4}
\end{equation}
The orbit space is represented by a certain 3D shape in the space of non-negative $(x, y, z)$,
sketched in Fig.~\ref{fig-S4-orbit}, left. 
Some of its features were already described in \cite{Degee:2012sk}. 
Here we finalize this description and 
use it to establish the necessary and sufficient BFB conditions. 
\begin{itemize}
	\item 
One notices the symmetry of this orbit space under the $x \leftrightarrow z$ exchange.\footnote{Ref.~\cite{Degee:2012sk} 
mentions this symmetry calling it ``unexpected'' and failing to identify the Higgs field transformation
behind the $x \leftrightarrow z$ exchange, which we describe in Appendix~\ref{appendix:S4.flat}.}
In fact, it displays an even stronger feature: its surface has flat directions parallel to the axis $x-z$.
That is, the full 3D orbit space is spanned between the two identical two-dimensional convex regions
on the faces $x = 0$ and $z=0$ which touch at a single point $x, z = 0$, $y = 1/3$.  
In Appendix~\ref{appendix:S4.flat} we prove this feature. 
	\item 
The back face of Fig.~\ref{fig-S4-orbit}, left, lies on the plane $x + y + z = 1$ 
and corresponds to the neutral orbit space. It has a trapezoidal shape, whose projection onto the 
$(x-z,y)$ plane is shown in Fig.~\ref{fig-S4-orbit}, right.
It extends along the $y$ direction up to $3/4$, and its vertices labeled as A, B, C, D correspond to the
following values of the variables and representative vev alignments:
\begin{eqnarray}
\mbox{point A:} &\quad& (x,y,z) = (0,0,1)\, \quad \Rightarrow \quad \lr{\phi_i^0} \propto (1, 0, 0),\nonumber\\
\mbox{point B:} &\quad& (x,y,z) = (1,0,0)\, \quad \Rightarrow \quad \lr{\phi_i^0} \propto (1, 1, 1),\nonumber\\
\mbox{point C:} &\quad& (x,y,z) = (1/4,\, 3/4,\, 0)\, \quad \Rightarrow \quad \lr{\phi_i^0} \propto (1,\omega,\omega^2),\nonumber\\
\mbox{point D:} &\quad& (x,y,z) = (0,\, 3/4,\, 1/4)\, \quad \Rightarrow \quad \lr{\phi_i^0} \propto (1, i, 0).
\label{points-S4-neutral}
\end{eqnarray}
The representative vev alignments are defined up to arbitrary sign flips and permutations of the doublets.
For example, point $D$ can be equally represented by $(1,\pm i,0)$, $(1,0,\pm i)$, and $(0,1,\pm i)$.
The vev alignments corresponding to the straight segments linking the vertices can be found in \cite{Degee:2012sk}.
	\item 
The charge-breaking part of the orbit space lies in the positive octant between the two planes:
\begin{equation}
1/4 \le x + y + z < 1\,.
\end{equation}
However, unlike in the case of rephasing invariant model, it does not fill this slab completely.
It contains certain arcs on the $x = 0$ and $z = 0$ planes, lying deep inside the charge-breaking part.
\begin{figure} [h]
	\centering
	\includegraphics[height=6cm]{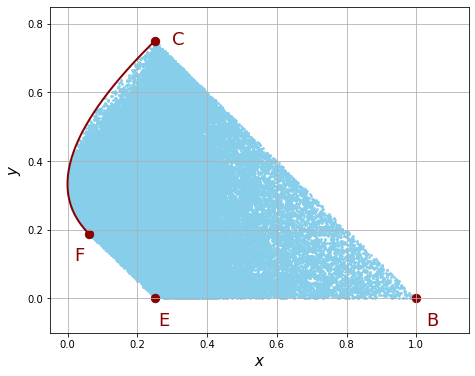}
	\caption{The orbit space of the $S_4$-symmetric 3HDM on the $z=0$ plane. 
		Shown are the corner points and the arc FC. The dots represent the numerical results of a random scan in the space of Higgs fields.}
	\label{fig-S4-x-y}
\end{figure}
In Fig.~\ref{fig-S4-x-y}, we show the structure of the orbit space on the face $z=0$.
Apart from the already familiar points B and C of the neutral orbit space, which intersects this
face along the line $x+y=1$, there are two other isolated points
inside the charge breaking orbit space: E at $(x,y) = (1/4, 0)$ and F at $(x,y) = (1/16, 3/16)$. 
All these points an consecutively linked by straight segments apart from the pair FC, which
is connected by the arc defined via
\begin{equation}
x = (1-\sqrt{3y})^2 \quad \mbox{for} \quad 3/16 \le y \le 3/4.\label{S4-arc}   
\end{equation}
This arc is derived in Appendix~\ref{appendix:S4.arc}.
Also shown in Fig.~\ref{fig-S4-x-y} is the outcome of a random scan over the Higgs field configurations 
satisfying $\phi_1^\dagger\phi_1 = \phi_2^\dagger\phi_2 = \phi_3^\dagger\phi_3$,
which ensures $z=0$. The points exactly cover the region outlined analytically, which serves as a numerical cross check. 
A similar picture takes place on the face $x=0$.
\end{itemize}

This geometric description makes it clear that the orbit space of the $S_4$ 3HDM 
does not possess any dangerous protrusions inside the charge-breaking orbit space
of the type shown in Fig.~\ref{fig-rho-z-2}.
This mere fact can be proven even without relying on the exact expression for the arcs.
In Appendix~\ref{appendix:S4.poly} we show that the entire orbit space
lies below the plane defined by $y = x+z+1/2$, which is tangent to the orbit space along the DC segment.

Therefore, applying the arguments of the previous section, we conclude:
if we work with the exact $S_4$ symmetry and insist on staying in a neutral vacuum, 
then the necessary and sufficient BFB conditions can be obtained
by checking positivity of $v_4$ only at the four vertices A, B, C, and D of the neutral orbit space:
\begin{equation}
\Lambda_0 + \Lambda_3 \ge 0\,, \quad \Lambda_0 + \Lambda_1 \ge 0\,, \quad 
\Lambda_0 + \frac{\Lambda_1 + 3\Lambda_2}{4} \ge 0\,, \quad \Lambda_0 + \frac{\Lambda_3 + 3\Lambda_2}{4} \ge 0\,. 
\label{BFB-S4-neutral}
\end{equation}
Recast in the notation of Eq.~\eqref{VA4-1}, these conditions read:
\begin{equation}
\lambda_1 \ge 0\,, \quad 
3\lambda_1 + \lambda_3 + \lambda_4 + \lambda_5 \ge 0\,,\quad 
3\lambda_1 + \lambda_3 + \lambda_4 -\frac{1}{2} \lambda_5 \ge 0\,,\quad 
4\lambda_1 + \lambda_3 + \lambda_4 - \lambda_5 \ge 0\,,
\label{BFB-S4-neutral-v2}
\end{equation}
or even more compactly 
\begin{equation}
\lambda_1 \ge 0\,, \quad 
3\lambda_1 + \lambda_3 + \lambda_4 + \mathrm{min}\left(\lambda_5,\, -\frac{\lambda_5}{2},\, \lambda_1 - \lambda_5\right) \ge 0\,.
\label{BFB-S4-neutral-v3}
\end{equation}
We stress that for the $S_4$ 3HDM, $\lambda_5$ can be of either sign.
In the case $\Lambda_1 = \Lambda_2$ (or, equivalently, $\lambda_5 = 0$) 
we recover the first two conditions of \eqref{BFB-U1U1} (for points A, B, and C),
after which the condition for point D is satisfied automatically.

Comparing the expressions above with the ones listed in subsection \ref{literature_conditions},
we see that \cite{Pramanick:2017wry} was the closest to deriving the full set of neutral BFB conditions for $S_4$.
It only misses the last constraint in Eq.~\eqref{BFB-S4-neutral-v2}, which was not stated explicitly but
which can be guessed from the extrapolation of their $A_4$ results to the $S_4$ case in the appendix.
In essence, the authors of \cite{Pramanick:2017wry} missed point D when deriving the BFB conditions for the $S_4$ case.

\subsection{Softly broken $S_4$}

As explained in the previous section, 
if the $S_4$ symmetry is softly broken, checking the positivity of $v_4$ only at neutral minima is no longer sufficient.
In this case, one needs to accompany Eq.~\eqref{BFB-S4-neutral} with the additional constraint
$v_4 \ge 0$ everywhere in the charge-breaking space.

Due to the symmetry of the orbit space and existence of flat directions,
it is sufficient to write these conditions on the faces $x=0$ and $z=0$.
Let us consider the plane $z=0$.
The condition $v_4 \ge 0$ at the two vertices E and F leads to two BFB conditions:
\begin{equation}
\mbox{point E:}\quad \Lambda_0 + \frac{\Lambda_1}{4} \ge 0\,, \qquad
\mbox{point F:} \quad \Lambda_0 + \frac{\Lambda_1 + 3\Lambda_2}{16} \ge 0\,.
\label{x-y-points-BFB-2}
\end{equation}
We stress again that they do not automatically follow from the conditions at points B and C,
since models with $\Lambda_0 < 0$ are allowed.
In the limit $\Lambda_1 = \Lambda_2$, which corresponds to the rephasing invariant model,
we recover expressions \eqref{final-U1U1-CB-Lam}.

In addition to the isolated points, 
requiring that $v_4 \ge 0$ everywhere along the arc leads to the following extra condition
together with its applicability range (see Appendix~\ref{appendix:S4.result} for the derivation):
\begin{equation}
\Lambda_0 + \frac{\Lambda_1\Lambda_2}{\Lambda_2 + 3\Lambda_1} \ge 0 \quad
\mbox{if}\quad \Lambda_1 > 0 \ \mbox{and} \ \Lambda_1 > |\Lambda_2|\,.\label{arc-condition-2} 
\end{equation}
For $|\Lambda_2| > \Lambda_1$ or for positive $\Lambda_1$, this new condition is not needed,
since the vertices take care of this case.

To summarize this discussion, we give the following final list of exact necessary and sufficient BFB conditions
for the $S_4$-symmetric 3HDM, which is applicable even to the softly broken case:
\begin{itemize}
	\item conditions \eqref{BFB-S4-neutral}, 
	\item
	conditions \eqref{x-y-points-BFB-2} and their counterparts with $\Lambda_1 \to \Lambda_3$;
	\item
	condition \eqref{arc-condition-2} and its counterpart with $\Lambda_1 \to \Lambda_3$.
\end{itemize}
If we deal with the case of an exact $S_4$ symmetry and select a neutral minimum, then the first item
is enough, as it provides the necessary and sufficient for this case.
If we study a softly broken $S_4$, then the last two items are imperative in order to avoid
pathological situations with a normally looking neutral minimum but a potential unbounded from below. 

\section{The $A_4$-symmetric case} \label{section-A4}

The orbit space of the $A_4$-symmetric 3HDM is represented by an elaborate shape
in the 4-dimensional space of variables $x,y,z,t$. 
The $(x,y,z)$-projection of this orbit space remains the same as in the $S_4$ 3HDM,
while the new variable $t$ is limited by $t^2 \le xy$.
The tricky point is that the equality is not reachable for all values of $(x,y,z)$,
which further complicates the shape of the orbit space.

Nevertheless, the neutral part of this orbit space was described in Ref.~\cite{Degee:2012sk}.
In particular, it was shown that a neutral vacuum can correspond either to one of the following isolated points
\begin{eqnarray}
\mbox{point A:} &\quad& (x,y,z,t) = (0,0,1,0)\, \quad \Rightarrow \quad \lr{\phi_i^0} \propto (1, 0, 0),\nonumber\\
\mbox{point B:} &\quad& (x,y,z,t) = (1,0,0,0)\, \quad \Rightarrow \quad \lr{\phi_i^0} \propto (1, 1, 1),\nonumber\\
\mbox{point C:} &\quad& (x,y,z,t) = (1/4,\, 3/4,\, 0,\, \sqrt{3}/4)\, \quad \Rightarrow \quad \lr{\phi_i^0} \propto (1,\omega^2,\omega),\nonumber\\
\mbox{point C${}^\prime$:} &\quad& (x,y,z,t) = (1/4,\, 3/4,\, 0,\, -\sqrt{3}/4)\, \quad \Rightarrow \quad \lr{\phi_i^0} \propto (1,\omega,\omega^2),
\label{points-A4-neutral-points}
\end{eqnarray}
or to a point on the circle 
\begin{equation}
\mbox{circle D:} \quad (x,y,z,t) = \left(\frac{3\cos^2\alpha}{4},\, \frac{3\sin^2\alpha}{4},\, \frac{1}{4},\,
\frac{3\sin\alpha\cos\alpha}{4}\right)\, \quad \Rightarrow \quad \lr{\phi_i^0} \propto (1, e^{i\alpha}, 0).
\end{equation}
Requiring that $v_4\ge 0$ at these points and everywhere on the circle leads to the following set of necessary conditions:
\begin{equation}
\Lambda_0 + \Lambda_3 \ge 0\,, \quad \Lambda_0 + \Lambda_1 \ge 0\,, \quad 
\Lambda_0 + \frac{\Lambda_1 + 3\Lambda_2 - \sqrt{3}|\Lambda_4|}{4} \ge 0\,,  
\label{BFB-A4-neutral-points}
\end{equation}
as well as
\begin{equation}
\Lambda_0 + \frac{\Lambda_3}{4} + \frac{3}{8}\left(\Lambda_1 + \Lambda_2 - \sqrt{(\Lambda_1-\Lambda_2)^2 + \Lambda_4^2}\right) \ge 0\,. 
\label{BFB-A4-neutral-circle}
\end{equation}
One can verify that, in the limit $\Lambda_4 \to 0$, this set of conditions is equivalent to \eqref{BFB-S4-neutral}.

In the parametrization of Eq.~\eqref{VA4-1}, these conditions can be written compactly as 
\begin{eqnarray}
\lambda_1 \ge 0\,, \quad 
3\lambda_1 + \lambda_3 + \lambda_4 + \lambda_5 \cos\left(\epsilon + \frac{2\pi k}{3}\right) \ge 0\,,\quad 
4\lambda_1 + \lambda_3 + \lambda_4 - \lambda_5 \ge 0\,,
\label{BFB-A4-neural-v2}
\end{eqnarray}
where the second condition must be checked for every $k = 0,1,2$.
In the limit $\epsilon = 0$ or $\pi$, we again recover the $S_4$ conditions \eqref{BFB-S4-neutral-v2}
valid for the neutral orbit space with positive or negative\footnote{Notice that in the $A_4$ case,
$\lambda_5$ is positive by definition; the negative option is taken care by the arbitrary phase $\epsilon$.} $\lambda_5$. 
Notice that the second condition here is stronger than the corresponding condition for the $S_4$ 3HDM.
None of the papers listed in section~\ref{literature_conditions} give exactly the same conditions for the neutral
directions in the $A_4$ model, with references \cite{Dekens:2011,Pramanick:2017wry} being the closest to this answer.

We {\em conjecture} that these constraints provide necessary and sufficient BFB conditions for the case of an exact $A_4$ symmetry.
In order to prove this conjecture, one needs to demonstrate that the charge-breaking part of the $A_4$ 3HDM orbit space
does not contain any protrusions, as described in connection with Fig.~\ref{fig-rho-z-2} above. 
An extensive numerical scan allowed us to visualize the 4D shape of this orbit space,
and its visual inspection lends support to the conjecture. However, at the moment, we do not have its analytic proof.

It goes without saying that, for the case of softly broken $A_4$, the above set of the BFB conditions becomes insufficient.
In this case, one would need to establish an explicit analytic description of the full 4D orbit
space and require that $v_4 \ge 0$ everywhere on its surface. This difficult task lies beyond the scope of the present work.

\section{Conclusions}

In summary, we demonstrated that several published expressions of the necessary and sufficient bounded-from-below conditions 
for the three-Higgs-doublet model with symmetry group $A_4$ are, at best, incomplete.
When revising these conditions, we came up with a remarkable phenomenon:
a valid set of necessary and sufficient BFB conditions for a multi-Higgs potential with an exact global symmetry group
may all of a sudden become insufficient if {\em soft} breaking terms are added.
We explained the origin of this phenomenon and showed how it should be avoided.
In particular, we derived for the first time the exact necessary and sufficient BFB conditions
valid for the (softly broken) $S_4$-symmetric 3HDM.

For the full $A_4$-symmetric case, we gave a set of conditions which we conjectured to be necessary and sufficient
in the case of exact $A_4$ symmetry, if we aim at a model with a neutral vacuum.
This full set of conditions did not appear before, although \cite{Dekens:2011} and \cite{Pramanick:2017wry} came closest.
Although we do yet not have the analytic proof that they are sufficient, 
this conjecture is supported by an extensive numerical scan in the 4D variable space.

However, these conditions will unavoidably become insufficient once soft breaking terms are added.
Unlike the $S_4$ symmetric case, we do not have the exact analytic description of the charge-breaking part of the full orbit space,
so we cannot present necessary and sufficient conditions which would be valid for the softly broken $A_4$.
Nevertheless, we believe these conditions should be expressible in terms of elementary functions.

\section*{Acknowledgments}
We thank Jo\~{a}o~P.\ Silva for useful discussions and comments on the paper. 
We acknowledge funding from the Portuguese \textit{Fun\-da\-\c{c}\~{a}o para a Ci\^{e}ncia e a Tecnologia} (FCT) through the projects UIDB/00777/2020 and UIDP/00777/2020, CERN/FIS-PAR/0004/2017, and PTDC/FIS-PAR/29436/2017, which are partially funded through Programa Operacional Competitividade e
Internacionalização (POCI), Quadro de Refer\^{e}ncia
Estrat\'{e}gica Nacional (QREN), and the European Union. 
I.P.I. also acknowledges the support from National Science Center, Poland, via the project Harmonia (UMO-2015/18/M/ST2/00518).


\appendix

\section{The bilinear formalism in 3HDM}\label{appendix:bilinear}

Some of the derivations below are done in the bilinear space of 3HDM
\cite{Nagel:PhD,Nishi:2006tg,Maniatis:2006fs,Nishi:2007nh,Ivanov:2010ww,Ivanov:2010wz,Maniatis:2014oza}.
For the sake of completeness, we give here the appropriate details using the notation of \cite{Ivanov:2010ww}.

The general renormalizable 3HDM Higgs potential is constructed from the gauge-invariant
bilinear combinations $\phi_i^\dagger \phi_j$, $i, j = 1,2,3$, which describe the gauge orbits in the Higgs space.\footnote{Strictly speaking, 
$\phi_i$ are operators acting on the Higgs Fock space.
However, for the purposes of checking stability and minimization of the potential,
one can view them as doublets of complex numbers.} 
It is convenient to group them in the following $1+8$ real bilinears:
\begin{equation}
r_0 = {1\over \sqrt{3}}\sum_i \phi_i^\dagger \phi_i\,,\quad r_a = \sum_{i,j} \phi_i^\dagger \lambda^a_{ij}\phi_j\,,
\quad a = 1, \dots, 8\,,
\label{rmu}
\end{equation}
where $\lambda^a$ are the generators of $SU(3)$.
The explicit expressions are
\begin{eqnarray}
&& r_0 = {(\phi_1^\dagger\phi_1) + (\phi_2^\dagger\phi_2) + (\phi_3^\dagger\phi_3)\over\sqrt{3}}\,,\ 
r_3 = {(\phi_1^\dagger\phi_1) - (\phi_2^\dagger\phi_2) \over 2}\,,\ 
r_8 = {(\phi_1^\dagger\phi_1) + (\phi_2^\dagger\phi_2) - 2(\phi_3^\dagger\phi_3) \over 2\sqrt{3}} \quad
\nonumber\\
&&r_1 = \Re(\phi_1^\dagger\phi_2)\,,\quad 
r_4 = \Re(\phi_3^\dagger\phi_1)\,,\quad 
r_6 = \Re(\phi_2^\dagger\phi_3)\,,\nonumber\\[2mm] 
&&r_2 = \Im(\phi_1^\dagger\phi_2)\,,\quad
r_5 = \Im(\phi_3^\dagger\phi_1)\,,\quad
r_7 = \Im(\phi_2^\dagger\phi_3)\,. \label{ri3HDM}
\end{eqnarray}
For future convenience, we also define the vector $n_a = r_a/r_0$.
The 3HDM orbit space in the $1+8$-dimensional real space of bilinears $(r_0,r_a)$ is defined by
\begin{equation}
r_0 \ge 0\,,\quad \vec n^2 \le 1\,,\quad \sqrt{3}d_{abc} n_a n_b n_c = {3 \vec n^2 - 1\over 2}\,,
\label{3HDMconditions}
\end{equation}
so that the modulus of the vector $\vec n$ is restricted by $1/4 \le \vec n^2 \le 1$ \cite{Ivanov:2010ww}.
Neutral vacua always lie on the surface of the outer cone $\vec n^2 = 1$,
while charge-breaking vacua lie strictly inside, $\vec n^2 < 1$. 

In the bilinear formalism, the Higgs potential becomes a quadratic form of $r_0$ and $r_a$:
\begin{equation}
\label{potential}
V = - M_0 r_0 - M_a r_a + {1 \over 2}\Lambda_{00} r_0^2 + \Lambda_{0a} r_0 r_a + {1 \over 2}\Lambda_{ab} r_a r_b\,,
\end{equation}
which is of help when dealing with certain problems.
The transition to the variables $x, y, z, t$ defined in Eq.~\eqref{VA4-3} further simplifies the structure of potential
making it a linear function of these variables.


\section{The BFB conditions for the rephasing invariant model: direct calculations}\label{appendix:checks}

Here we will consider $\lambda_5=0$ and thus the last term in \eqref{VA4-1} vanishes. Then we are left with a $U(1)\times U(1)$ symmetric potential with an extra permutation symmetry. The necessary and sufficient BFB conditions for a rephasing invariant potential were derived in \cite{Faro:2019vcd}. The idea is to write the potential as a quadratic form of independent non-negative variables and apply the \textit{copositivity} of the matrix associated with the quadratic form \cite{Kannike:2012pe}. For neutral directions one sees clearly that it is possible to find a parametrization where the potential does not depend on the doublets' phases, and $V_N$ defined below is immediately written as a quadratic form. 
This does not immediately hold for charge breaking directions, thus one needs first to find the directions that minimize the potential, for those directions write the potential as a quadratic form, and finally apply the copositivity conditions.
To this end we first rewrite the quartic part of Eq.~\eqref{VA4-1} as $V_N + V_{CB}$, where
\begin{eqnarray}
    V_N &=& \frac{a}{2}\left[(\phi_1^\dagger\phi_1)^2 + (\phi_2^\dagger\phi_2)^2 + (\phi_3^\dagger\phi_3)^2\right] + b\left[(\phi_1^\dagger\phi_1)(\phi_2^\dagger\phi_2) + (\phi_1^\dagger\phi_1)(\phi_3^\dagger\phi_3)+ (\phi_2^\dagger\phi_2)(\phi_3^\dagger\phi_3)\right]\nonumber\\
    V_{CB} &=& c\left[z_{12} + z_{13}+z_{23}\right], \quad \textrm{with} \quad z_{ij} = (\phi_i^\dagger\phi_i)(\phi_j^\dagger\phi_j) - (\phi_i^\dagger\phi_j)(\phi_j^\dagger\phi_i) \geq 0\,.
    \label{CB_part}
\end{eqnarray}
Because the doublet bilinears $\phi_i^\dagger \phi_j$ are invariant under the SM gauge group, 
we can use a suitable basis and parametrize doublets as follows:
\begin{eqnarray}
\phi_1= \sqrt{r_1}\doublet{0}{1}, \quad 
\phi_2= \sqrt{r_2}\doublet{\sin{(\alpha_2)}}{\cos{(\alpha_2)}e^{i\beta_2}}, \quad
\phi_3= \sqrt{r_3}e^{i\gamma}\doublet{\sin{(\alpha_3)}}{\cos{(\alpha_3)}e^{i\beta_3}}
\label{eq:doublet_par}
\end{eqnarray}
These $r_i$ are not to be confused with the vector $r_a$ from the previous appendix.
With this parametrization, we can write the neutral part of the potential as:
\begin{equation}
    V_N = \frac{a}{2}(r_1^2 + r_2^2 + r_3^2) + b(r_1r_2 + r_1r_3 + r_2r_3) \equiv \frac{1}{2}A_{ij}r^i r^j
    \label{eq:VN_par}
\end{equation}
This quadratic form is non-negative definite in the first octant of its variables if and only if the entries of matrix $A$ 
satisfy the copositivity conditions \cite{Kannike:2012pe}. In our case, due to the simple form of the matrix, they are
\begin{equation}
a \geq 0\,,\quad b \geq -\frac{a}{2}\,. \label{eq:st_1}
\end{equation}
However, we must also check charge breaking directions to see if stronger conditions are required. 
The charge breaking part of the potential written with parametrization \eqref{eq:doublet_par} is:
\begin{eqnarray}
   V_{CB} &=& c \big\{ r_1 r_2 \sin^2{(\alpha_2)} + r_1 r_3 \sin^2{(\alpha_3)} + 
    r_2 r_3 \left[1 - \sin^2{(\alpha_2)} \sin^2{(\alpha_3)} - \cos^2{(\alpha_2)}\cos^2{(\alpha_3)}\right] \nonumber\\[2mm] 
    && - \frac{1}{2}r_2 r_3 \sin{(2\alpha_2)}\sin{(2\alpha_3)}\cos{(\beta_2-\beta_3)}\big\}\,.
    \label{eq:cb_potential}
\end{eqnarray}
Minimizing equation \eqref{eq:cb_potential} with respect to the angular variables, one arrives to the following equalities:
\begin{equation}
    \frac{\sin{(2(\alpha_2 \mp \alpha_3))}}{r_1} = \frac{\sin{(2\alpha_3)}}{r_2} = \frac{\sin{(-2\alpha_2)}}{r_3}
    \label{eq:CB_equalities}
\end{equation}
First, we can look into trivial solutions of \eqref{eq:CB_equalities} when all sines are equal to zero.
These are obtained when $\alpha_2$ and $\alpha_3$ are multiples 
of $\pi/2$. In this case the potential simplifies to one of the three possible forms:
\begin{eqnarray}
c(r_1 r_2 + r_1 r_3), \quad c(r_1 r_2 + r_2 r_3), \quad c(r_1 r_3 + r_2 r_3) 
\label{eq:triv_directions}
\end{eqnarray}
Applying now the copositivity conditions to the quadratic form of $V_N + V_{CB}$ with the above expression for $V_{CB}$
produces the following new relations:
\begin{equation}
\textrm{if}\quad -\frac{a}{2}\leq b \leq a:  \quad b + c\geq -\sqrt{\frac{a(a+b)}{2}}\,, \qquad
\textrm{if}\quad b \geq a: \quad b + c\geq -a\,. \label{eq:st_2}
\end{equation}
Finally, it is necessary to check non-trivial solutions of Eq.~\eqref{eq:CB_equalities}.
For that, we write it as the law of sines, just like in \cite{Faro:2019vcd}, and get the following simple form of the charge breaking potential:
\begin{equation}
    V_{CB} = \frac{c}{4}\left(r_1+r_2+r_3\right)^2\,.
\end{equation}
This expression is applicable within the open tetrahedron described in \cite{Faro:2019vcd}. 
Writing the copositivity conditions for $V_N + V_{CB}$, we arrive at the following additional relation:
\begin{equation}
b+\frac{c}{2}\geq -(a+b+c) \label{eq:st_3}
\end{equation}
Ultimately, combining the conditions we have obtained at all three steps and keeping only the stronger ones, 
we get the following simple necessary and sufficient conditions for our potential to be bounded from below:
\begin{align}
a \geq 0,\quad
b \geq - \frac{a}{2} \label{eq:st_final_b},\\
\textrm{If} \quad -\frac{a}{2}\leq b \leq a: \quad b+\frac{c}{2}\geq -(a+b+c) \label{eq:st_final_c},\\
\textrm{If} \quad b > a: \quad b + c\geq -a \label{eq:st_final_d}
\end{align}
We see that, for $c < 0$, it is mandatory to check charge breaking directions that deliver conditions \eqref{eq:st_final_c} and \eqref{eq:st_final_d}. 
Notice also that the last two conditions are not conflicting but complementary. Thus, one can also impose both of them simultaneously,
without the need to verify whether $b> a$ or $b < a$.

Finally, linking the original potential \eqref{VA4-1} with $\lambda_5=0$ with $V_N+V_{CB}$ via 
$a = 2\lambda_1$, $b = 2\lambda_1 + \lambda_3 + \lambda_4$, $c = - \lambda_4$, 
we get the final set of necessary and sufficient BFB conditions for the rephasing invariant version of the $A_4$ 3HDM:
\begin{eqnarray}
\mbox{neutral:}&\qquad& \lambda_1 \ge 0\,, \quad
3\lambda_1 + \lambda_3 + \lambda_4 \ge 0\,, \label{final-U1U1-neutral-app}\\
\mbox{charge-breaking:}&\qquad& 4\lambda_1 + \lambda_3 \ge 0\,, \quad
3\lambda_1 + \lambda_3 + {1\over 4}\lambda_4 \ge 0\,.\label{final-U1U1-CB-app}
\end{eqnarray}


\section{The orbit space of the $S_4$ 3HDM}\label{appendix:S4}

\subsection{Flat directions}\label{appendix:S4.flat}

An important feature of the orbit space is that it is flat along the coordinate $x-z$.
Namely, if a point with given $x_0$, $y_0$, and $z_0$ belongs to the orbit space, then
the entire straight segment 
\begin{equation}
x = \alpha(x_0 +z_0)\,,\quad y = y_0\,,\quad z= (1-\alpha)(x_0 +z_0)\quad \mbox{with} \quad \alpha \in [0,1]
\end{equation}
also belongs to the orbit space. 

The proof relies on the expressions for the bilinears \eqref{ri3HDM}.
Let us perform an $SO(3)$ rotation of the three Higgs doublets $\phi_i$.
It induces an $SO(8)$ rotation of the bilinears $r_a$ which acts,
separately, within the subspaces of purely real symmetric and purely imaginary antisymmetric generators.
As a result, one obtains an orthogonal rotation in the 3D subspace $V_A = (n_2,n_5,n_7)$
and an orthogonal rotation in the 5D subspace $V_S = (n_1,n_3,n_4,n_6,n_8)$.
Consequently, this rotation leaves invariant $y$ and $x+z$ but it changes $x-z$.

Next, we need to prove that, starting from any $x_0$ and $z_0$,
one can apply such orthogonal transformations to go all the way to the boundaries, 
that is, to set $x=x_0 + z_0$ with $z = 0$ and $x = 0$ with $z=x_0 + z_0$.
Consider 
\begin{equation}
K_{ij} = \sum_{a=1,3,4,6,8} n_a \lambda^a_{ij}\,,
\end{equation}
where we used only the subspace $V_S$.
$K$ is a real symmetric traceless $3\times 3$ matrix, 
transforming as a 5-plet of $SO(3)$.
This matrix can be diagonalized by an orthogonal transformation.
Since the diagonal generators of $SU(3)$ are $\lambda_3$ and $\lambda_8$, 
this brings the vector $n_a$ within this subspace to the $(n_3, n_8)$.
Thus, we arrive at $x = 0$ and $z$ taking its maximal value.

Alternatively, we can perform another orthogonal transformation 
to make all $\phi_i^\dagger \phi_i$ equal.
This transformation populates the off-diagonal entries of $K$ setting its diagonal entries to zero.
In this way, the vector $n$ is brought to the subspace $(n_1, n_4, n_6)$.
We arrive at $z=0$ and a maximal possible $x$.

One consequence of the property just proved is the $x \leftrightarrow z$ symmetry
of the orbit space seen in Fig.~\ref{fig-S4-orbit}.
Notice that this is {\em not} a symmetry of the model, at least, when $\Lambda_1 \not = \Lambda_3$.
But it is a symmetry of its orbit space.

Another property is that, when constructing the exact shape of the orbit space,
it is enough to consider only one flat face, for example $z = 0$.
Once we know its exact shape on this face, we reconstruct the entire orbit space.
It also has important consequences for the geometric derivation of the BFB conditions. 
The flatness of the orbit space along the axis $x-z$ allows us to write down the BFB conditions 
only for the two faces with $x=0$ and $z=0$.
Once the BFB conditions at these two faces are satisfied, they are automatically satisfied for the 
entire orbit space.

\subsection{Circumscribed polyhedron}\label{appendix:S4.poly}

Before describing the exact shape of the orbit space, let us show that
it satisfies yet another linear inequality:
\begin{equation}
y \le x+z+1/2\,.\label{S4-fourth-plane-again} 
\end{equation}
This plane is tangent to the orbit space along the segment CD and it descents into the charge-breaking part
with the slope which matches the local orbit space slope.

Indeed, explicit calculations give
\begin{equation}
x-y+z+{1\over 2} = {3 \over 2}\,\frac{(\phi_1^\dagger\phi_1)^2+(\phi_2^\dagger\phi_2)^2+(\phi_3^\dagger\phi_3)^2 + 
	[(\phi_1^\dagger\phi_2)^2 + (\phi_2^\dagger\phi_3)^2 + (\phi_3^\dagger\phi_1)^2 + H.c.]}{(\phi_1^\dagger\phi_1+\phi_2^\dagger\phi_2+\phi_3^\dagger\phi_3)^2}
\label{Qab-1}
\end{equation}
Let us introduce the following $(2,0)$ tensor of the electroweak $SU(2)$ already used in \cite{Degee:2012sk}:
\begin{equation}
Q_{\alpha\beta} = (\phi_1)_\alpha (\phi_1)_\beta + (\phi_2)_\alpha (\phi_2)_\beta + (\phi_3)_\alpha (\phi_3)_\beta \,,
\end{equation}
where indices $\alpha, \beta$ are electroweak indices within each doublet.
Then, the numerator of \eqref{Qab-1} is elegantly written as the norm of this tensor, which of course must be non-negative:
\begin{equation}
Q_{\alpha\beta}^\dagger Q_{\alpha\beta} = |Q|^2 \ge 0\,.
\end{equation}
The equality holds only when the entire tensor $Q_{\alpha\beta} = 0$, which is possible only when all the doublets 
$\phi_i$ are proportional to one another (that is, we are in the neutral orbit space).
This proves \eqref{S4-fourth-plane-again}.

The consequence of this fact is the following.
If we study the exact $S_4$ symmetric potential and if we care only about models with neutral minima,
then it is safe to write the BFB conditions only for the four vertices of the neutral orbit space.
The property just proved will guarantee that in that case there is no dangerous protrusion coming from 
the charge-breaking orbit space.

However, if we want to extend the results to the more phenomenologically viable softly broken $S_4$,
then these conditions are no longer sufficient, and we will need to include an extra set of conditions
coming from the charge-breaking orbit space.

\subsection{The arc}\label{appendix:S4.arc}

\subsubsection{Formulation of the problem}

Let us consider the orbit space on the plane $z=0$.
It corresponds to the situation when all three doublets have equal norms.
We can take this common norm out, writing the doublets as
$\phi_i = |\phi_i| \hat\phi_i$, where the rescaled doublets $\hat\phi_i$ have unit norm.
Writing down the two relevant variable $x$ and $y$, we prefer to combine them as 
\begin{eqnarray}
x+y &=& {1 \over 3}\left(|\hat\phi_1^\dagger\hat\phi_2|^2 + |\hat\phi_2^\dagger\hat\phi_3|^2 + |\hat\phi_3^\dagger\hat\phi_1|^2\right)\,,\nonumber\\[2mm]
x-y &=& {1 \over 6}\left[(\hat\phi_1^\dagger\hat\phi_2)^2 + (\hat\phi_2^\dagger\hat\phi_3)^2 + (\hat\phi_3^\dagger\hat\phi_1)^2 + H.c.\right]\,.
\end{eqnarray}
We see that these two variables have very similar form.
In particular, by selecting all doublets real, one sets $y=0$ so that $x-y$ can be as large as $x+y$.
But what is the {\em minimal} value of $x-y$ for a given $x+y$?

Let us define, for brevity, the following variables:
$(\hat\phi_i^\dagger\hat\phi_j)^2 \equiv p_{ij}\,e^{i\psi_{ij}}$.
Then
\begin{eqnarray}
x+y &=& {1 \over 3}(p_{12}+p_{23}+p_{31})\,, \nonumber\\[2mm]
x-y &=& {1 \over 3}(p_{12}\cos\psi_{12} + p_{23}\cos\psi_{23} + p_{31}\cos\psi_{31})\,. 
\end{eqnarray}
It is important to keep in mind that the six new variables $p_{ij}$ and $\psi_{ij}$
are not independent. Let us denote $\psi_{12}+\psi_{23}+\psi_{31} \equiv \Psi$ 
and $p_{12}+p_{23}+p_{31} \equiv P$.
Then, the value of $\Psi$ depends on $p_{ij}$ as derived in \cite{Faro:2019vcd}:
\begin{equation}
\cos^2(\Psi/2) = \frac{(P - 1)^2}{4p_{12}p_{23}p_{31}}\,.
\label{relation-Faro}
\end{equation}
Our task is, for a given $P = 3(x+y)$, find the minimal value of $x-y$ by varying individual $p_{ij}$
and $\psi_{ij}$ and keeping in mind the above relation.

First, since Eq.~\eqref{relation-Faro} is bounded by 1, arbitrarily small $P$'s are not allowed.
For a given $P$, the smallest value for this expression is obtained 
under the assumption of equipartition $p_{12} = p_{23} = p_{31} = P/3$ and is equal to
\begin{equation}
\cos^2(\Psi/2) = \frac{27(P - 1)^2}{4 P^3}\,.
\label{relation-Faro-equal}
\end{equation}
Requiring it not to exceed 1 places a lower bound on $P \ge 3/4$.
This translates to the already familiar constraint $x+y \ge 1/4$.

\subsubsection{The main solution}

Now, in order to minimize $x-y$ for a given $P$, let us explicitly include the constraints:
\begin{equation}
x-y = {1 \over 3}\left[p_{12}\cos\psi_{12} + p_{23}\cos\psi_{23} + (P - p_{12} - p_{23})\cos(\Psi-\psi_{12}-\psi_{23})\right]\,,
\label{x-y-ready}
\end{equation} 
where $\Psi$ depends on $p_{12}$ and $p_{23}$ as
\begin{equation}
\cos^2(\Psi/2) = \frac{(P - 1)^2}{4p_{12}p_{23}(P - p_{12} - p_{23})}\,.
\label{relation-Faro-2}
\end{equation}
Next, differentiating \eqref{x-y-ready} with respect to $\psi_{12}$ and $\psi_{23}$ allows us to deduce the following important equalities:
\begin{equation}
p_{12}\sin \psi_{12} = p_{23}\sin \psi_{23} = (P - p_{12} - p_{23})\sin(\Psi-\psi_{12}-\psi_{23})\,.\label{sine-relation}
\end{equation}
One possibility is that all sines are equal to zero. We call this special case the trivial solution, which we will analyze later, 
and now we proceed with the non-trivial solution of sines being non-zero.
Differentiating \eqref{x-y-ready} with respect to $p_{12}$ and $p_{23}$ and working out the expressions, one arrives to the simple equation:
\begin{equation}
\sin\left(\frac{\Psi}{2}-\psi_{12}\right) = \sin\left(\frac{\Psi}{2}-\psi_{23}\right)\,.
\end{equation}
The main solution is $\psi_{12} = \psi_{23}$, which also implies $p_{12} = p_{23}$. 
The other possibility $\Psi/2 - \psi_{12} = \pi - (\Psi/2 - \psi_{23})$
would result in $\Psi - \psi_{12} - \psi_{23} = \pi$ and lead to the trivial solution of \eqref{sine-relation}.

Now, equation \eqref{sine-relation} can be rewritten as  
$(P - 2p_{12})\sin(\Psi-2\psi_{12}) = p_{12}\sin \psi_{12}$. 
Inserting it into the derivative of \eqref{x-y-ready} with respect to $p_{12}$ and simplifying the resulting expression one gets the equality:
\begin{equation}
\sin\left(\frac{\Psi}{2}-\psi_{12}\right) = \sin\left(2\psi_{12} - \frac{\Psi}{2}\right)\,.
\end{equation}
First, let us notice that the solution of this equation
when the two angles sum up to $\pi$ again leads to the trivial case for the sine relations \eqref{sine-relation}.
Thus, we consider the main solution when the two angles are equal. It leads to
\begin{equation}
\psi_{12} = \psi_{23} = \frac{\Psi + 2\pi k}{3}\,, \quad p_{12} = p_{23} = \frac{P}{3}\,.\label{minimum-points}
\end{equation}
Thus, we obtain the situation of equipartition, 
and the relation between $\Psi$ and $P$ is given by \eqref{relation-Faro-equal}.
The value of $x-y$ is
\begin{equation}
x-y = (x+y)\cos\left(\frac{\Psi}{3} + \frac{2\pi k}{3}\right)\,.
\end{equation}
For $\Psi \in [0, 2\pi]$, the minimal value is given by $k=1$ and can be written as
\begin{equation}
(x-y)_{min} = -(x+y)\cos\left(\frac{\Psi - \pi}{3}\right)\,.\label{x-y-min}
\end{equation}
The final step is to obtain a single algebraic equation relating $x-y$
and $x+y$ and describing the arc. 
Linking \eqref{relation-Faro-equal} with \eqref{x-y-min} through the formula for the cosine of a triple angle,
one can arrive at a single cubic relation between $x$ and $y$, which, after some algebra, simplifies to
the very compact relation:
\begin{equation}
\sqrt{3y} = 1\pm\sqrt{x}\,, \quad \mbox{or}\quad x = (1-\sqrt{3y})^2\,.\label{ridiculous}
\end{equation}
The upper branch extends to $x = 1/4$, while the lower branch terminates at $x = 1/16$.

\subsubsection{The ``trivial'' solution}

Let us now consider the trivial solution, namely, when all sines in Eq.~\eqref{sine-relation} are equal to zero.
This is possible only when all the angles including $\Psi$ are multiples of $\pi$.
If $\Psi = \pi$, then $\cos\Psi/2 = 0$, and Eq.~\eqref{relation-Faro} tells us that this can happen only at $P = 1$.
Only at this isolated point can all three angles be equal to $\pi$, so that 
$x-y$ can become $-(x+y)$. This case, however, is not new and is already covered by \eqref{x-y-min}.

Less trivial is the situation of $\Psi = 0$ and $\psi_{12}=\psi_{23} = \pi$.
For this configuration
\begin{equation}
x-y = -{1\over 3}(2p_{12}+2p_{23} - P)\quad\mbox{with}\quad 4p_{12}p_{23}(P-p_{12}-p_{23}) = (P-1)^2\,.
\end{equation}
We need to find the minimal value of this function for a given $P = 3(x+y)$ and
check if it lies deeper than the non-trivial solution \eqref{x-y-min}.

By applying the Lagrange multiplier technique for constrained extremization, 
we can deduce that $p_{12} = p_{23}$.
Next, we solve 
\begin{equation}
4 p_{12}^2(P-2p_{12}) = (P-1)^2
\end{equation} 
for $p_{12}$. To do it for this cubic equation, we first solve for $P$ as a function of $p_{12}$
and then invert the dependence. Thus, one gets two solutions, the first leads to:
\begin{equation}
x-y = -{1\over 3}(P-2)\,.
\end{equation}
This minimum is always less deep than the non-trivial solution found previously:
\begin{equation}
P\cos\left(\frac{\Psi - \pi}{3}\right) \ge {P\over 2} > P-2 \quad \mbox{for}\quad P\le 3\,.
\end{equation}
Thus, this solution can be disregarded. The second solution leads to
\begin{equation}
p_{12} = {1\over 4} + {1\over 2}\sqrt{P - 3/4} \quad \Rightarrow \quad
x-y = -{1\over 3}\left(1 - P + 2\sqrt{P - 3/4}\right)\,.
\end{equation}
We kept here only one of the two solutions for $p_{12}$ since the other leads to a higher minimum. 
Finally, one can find that
\begin{equation}
P\cos\left(\frac{\Psi - \pi}{3}\right) \ge 1 - P + 2\sqrt{P - 3/4}\,. 
\end{equation}
The equality holds at the special point $P = 1$. 
The summary is that the ``trivial'' solutions to the sine relations \eqref{sine-relation}
never produce points which lie deeper than the non-trivial solutions.
Thus, they can be disregarded when plotting the shape of the orbit space.

\subsection{The net result: description of the $(x,y)$ plane}\label{appendix:S4.result}

This allows us to fully construct the orbit space on the $(x,y)$ plane as depicted in Fig. \ref{fig-S4-orbit}.
One observes the already familiar points B and C in the neutral orbit space
and two vertices deep inside the charge breaking orbit space: 
E at $(x,y) = (1/4, 0)$ and F at $(x,y) = (1/16, 3/16)$. 
The BFB conditions at these points are:
\begin{eqnarray}
&&\mbox{point B:}\quad \Lambda_0 + \Lambda_1 \ge 0\,, \qquad 
\mbox{point C:}\quad \Lambda_0 + \frac{\Lambda_1 + 3\Lambda_2}{4} \ge 0\nonumber\\[2mm]
&&\mbox{point E:}\quad \Lambda_0 + \frac{\Lambda_1}{4} \ge 0\,, \qquad
\mbox{point F:} \quad \Lambda_0 + \frac{\Lambda_1 + 3\Lambda_2}{16} \ge 0\,.
\label{x-y-points-BFB}
\end{eqnarray}
All of them are linked by straight segments apart
from the arc FC which is given by the equation $x = (1 -\sqrt{3y})^2$ defined for
$3/16 < y < 3/4$.
This arc represents a convex part of the orbit space, therefore,
we need to require the BFB conditions for all points on this arc:
\begin{equation}
\Lambda_0 + \Lambda_1 (1 -\sqrt{3y})^2 + \Lambda_2 y > 0 \quad \forall y \in (3/16, 3/4)\,.
\end{equation}
To reduce it to a set of inequalities, we need to check whether this function
possesses a minimum anywhere inside the $y$ domain. 
If it does not, then the BFB conditions at the endpoints C and F are sufficient.
If it does, we require this minimum value to be non-negative.
A straightforward analysis shows that an extra condition emerging from the arc is
\begin{equation}
\Lambda_0 + \frac{\Lambda_1\Lambda_2}{\Lambda_2 + 3\Lambda_1} \ge 0 \quad
\mbox{if}\quad \Lambda_1 > 0 \ \mbox{and} \ \Lambda_1 > |\Lambda_2|\,.\label{arc-condition} 
\end{equation}
This condition complements Eqs.~\eqref{x-y-points-BFB}.

\end{document}